\documentclass[manuscript]{aastex6}
\newcommand{\wind}{\textit{Wind}}
\newcommand{\ace}{\textit{ACE}}
\newcommand{\ulysses}{\textit{Ulysses}}
\newcommand{\stereo}{\textit{STEREO}}
\newcommand{\goes}{\textit{GOES 12}}
\newcommand{\soho}{\textit{SOHO${/}$}LASCO}
\newcommand{\insitu}{\textit{in situ}}
\newcommand{\kms}{km s$^{-1}$}
\usepackage{amsmath}
\shorttitle{Propagation of 2005 May 6 and 13 CMEs}
\shortauthors{Zhao et al.}
\begin{document}
\title{Propagation Characteristics of Two Coronal Mass Ejections
From the Sun Far into Interplanetary Space}
\author{Xiaowei Zhao\altaffilmark{1,2}, Ying D. Liu\altaffilmark{1,2},
        Huidong Hu\altaffilmark{1,2}, and Rui Wang\altaffilmark{1}}
\altaffiltext{1}{State Key Laboratory of Space Weather,
        National Space Science Center,
        Chinese Academy of Sciences, Beijing 100190, China;
        \href{mailto:liuxying@spaceweather.ac.cn}{liuxying@spaceweather.ac.cn}}
\altaffiltext{2}{University of Chinese Academy of Sciences, Beijing 100049, China}

\begin{abstract}
Propagation of coronal mass ejections (CMEs) from the Sun far into interplanetary space is not well understood due to limited observations. In this study we examine the propagation characteristics of two geo-effective CMEs, which occurred on 2005 May 6 and 13, respectively. Significant heliospheric consequences associated with the two CMEs are observed, including interplanetary CMEs (ICMEs) at the Earth and \ulysses, interplanetary shocks, a long-duration type II radio burst, and intense geomagnetic storms. We use coronagraph observations from \soho, frequency drift of the long-duration type II burst, \insitu{} measurements at the Earth and \ulysses, and magnetohydrodynamic (MHD) propagation of the observed solar wind disturbances at 1 AU to track the CMEs from the Sun far into interplanetary space. We find that both of the two CMEs underwent a major deceleration within 1 AU and thereafter a gradual deceleration when they propagated from the Earth to deep interplanetary space due to interactions with the ambient solar wind. The results also reveal that the two CMEs interacted with each other in the distant interplanetary space even though their launch times on the Sun were well separated. The intense geomagnetic storm for each case was caused by the southward magnetic fields ahead of the CME, stressing the critical role of the sheath region in geomagnetic storm generation, although for the first case there is a corotating interaction region involved.
\end{abstract}


\section{Introduction}\label{Intro}

Coronal mass ejections (CMEs) are large eruptions of plasma and magnetic fields from the solar corona and can cause severe space weather effects. Interplanetary CMEs (ICMEs) are regarded as the heliospheric counterpart of CMEs. ICMEs are often associated with interplanetary shocks and prolonged southward components of magnetic fields. A southward field component can reconnect with the dayside geomagnetic fields and trigger geomagnetic storms \citep{1961Dungey, 1994Gonzalez, 1997Tsurutani}. Although the source regions of CMEs and their propagation within 1 AU have been extensively studied, their evolution in the deep interplanetary space is not well understood due to the limitation of observations. Multi-spacecraft remote sensing and \insitu{} observations are crucial for understanding the propagation of CMEs in interplanetary space.    

Previous studies indicate that different types of CMEs may have different propagation histories. Fast CMEs may undergo a deceleration while slow CMEs are accelerated due to interactions with the ambient solar wind \citep{1999Sheeley, 2000Gopalswamy}. Empirical models have been proposed to investigate the propagation of CMEs within 1 AU combining coronagraph observations and \insitu{} measurements \citep{1999Lindsay,2001GopalswamyL}. \citet{2001GopalswamyL} suggest that CMEs experience a deceleration out to 0.76 AU and then they move with a constant speed. These studies are based on coronagraph observations, whose field of view (FOV) is limited to 30 R$_\sun$. A propagation model has been proposed using the frequency drift of interplanetary type II radio bursts to invert the distance of CME-driven shocks \citep{2007Reiner, 2008LiuL}. Based on a statistical analysis of interplanetary type II radio bursts, \citet{2007Reiner} argue that the deceleration cessation distances of CMEs range from 0.3 AU to beyond 1 AU. Doppler scintillation measurements of shock locations and speeds indicate that fast shocks may experience substantial deceleration near the Sun \citep{1985Woo, 1988Woo}. \citet{1986Cane} suggest that fast shocks may decelerate rapidly while slow shocks propagate at about constant speed, based on the comparison between the result of \citet{1985Woo} and \insitu{} measurements at Helios. Combining coronagragh observations, type II radio bursts, interplanetary scintillation technique (IPS) and \insitu{} observations at \wind, \citet{2009Gonzalez} argue that CMEs/shocks undergo a gradual deceleration within 1 AU. A general picture of Sun-to-Earth propagation of fast CMEs (with speeds above 1000 \kms ) has been found by \citet{2013Liu} combining stereoscoptic wide-angle heliocentric imaging observations from \stereo, interplanetary type II radio bursts, and \insitu{} measurements at 1 AU. They reveal that fast CMEs experience an impulsive acceleration, then decelerate in a relatively short timescale, and finally move with a roughly constant speed or a gradual deceleration. \citet{2016Liu} further provide the propagation characteristics of slow CMEs (with speeds below 400 \kms) in the Sun-Earth space. They find that slow CMEs are gradually accelerated to the ambient solar wind speed around 20--30 solar radii and then comove with the solar wind. 

Interplanetary propagation of CMEs can be influenced by their interactions with solar wind structures, including other CMEs \citep[e.g.,][]{2001GopalswamyY, 2012Liu, 2014LiuL, 2012Lugaz, 2012Temmer}, which will make their propagation more complex. \citet{2012Temmer} find that a fast CME can be significantly slowed down by interacting with another CME. A CME can also be deflected by interacting with other CMEs in the corona and interplanetary space \citep{2001GopalswamyY, 2012Lugaz, 2014LiuL}. \citet{2014LiuL} suggest that an earlier CME can remove some of the solar wind plasma and give rise to a modest deceleration of the later CME, which they call ``preconditioning''. The importance of the preconditioning of the upstream solar wind for CME propagation has been confirmed by \citet{2015Temmer} and \citet{2015Cash}. Other solar wind structures, such as high speed streams, can change CME propagation in a similar way. For example, slow CMEs can be accelerated by high speed streams from behind \citep{2015Liu, 2016Liu, 2015Kataoka}.

CME kinematics can be obtained by combining multi-spacecraft remote sensing and \insitu{} observations with different modeling techniques. \citet{2006Thernisien} have proposed a Graduated Cylindrical Shell (GCS) model to reconstruct CMEs. The GCS model can acquire the heliocentric distances of CME leading front and has been successfully applied to \textit{SOHO} and \stereo{} imaging observations \citep[e.g.,][]{2009Thernisien, 2010Liu, 2013Cheng, 2015Mishra}. The frequency drift of type II radio bursts produced by CME-driven shocks can be converted to radial distances using a proper solar wind density model \citep[e.g.,][]{2007Reiner, 2008LiuL, 2013Liu, 2015Cremades, 2016Hu}. Using the electron density model of \citet[][referred to as the Leblanc density model hereafter]{1998Leblanc}, \citet{2013Liu} obtain the radial distances of CME-driven shocks from type II burst observations and find general consistency with those derived from triangulation techniques based on \stereo{} imaging observations. \citet{2008LiuL} have investigated the propagation of a CME/shock from the Sun far into interplanetary space combining the frequency drift of type II radio emissions, \insitu{} observations at 1 AU and \ulysses, and a magnetohydrodynamic (MHD) model. \citet{2014Richardson} compares the speeds of 11 ICMEs at the Earth and \ulysses{} and suggests that ICMEs with speeds above the solar wind at the Earth continue to decelerate out to \ulysses{} though at a slower rate than that in the Sun-Earth space. Based on the derived CME/shock kinematics, we can estimate their arrival times at the Earth and investigate their interplanetary propagation characteristics. 

The main purpose of this paper is to investigate the propagation and evolution of two geo-effective CMEs in the heliosphere, which occured on 2005 May 6 and 13, respectively. The two CMEs are well observed by a fleet of spacecraft, including coronagraph observations from \soho, \insitu{} measurements at the Earth and \ulysses, and a long duration type II burst from \wind. This provides a great opportunity to study the propagation of CMEs from the Sun to deep interplanetary space. We combine these different data sets with modeling techniques to give a comprehensive view of the two geo-effective CMEs. Although there are some previous studies looking at the 2005 May 13 CME individually \citep[e.g.,][]{2006Yurchyshyn, 2007Reiner,2009Xie,2010Bisi,2010Gopalswamy, 2014Manchester}, we study the two geo-effective CMEs together and pay particular attention to the propagation characteristics of the two CMEs from the Sun far into interplanetary space. It is also interesting to show that, although the launch times of the two CMEs are well separated, they can still interact with each other in interplanetary space. We illustrate how the arrival time of a CME/shock can be estimated at the Earth using merged remote-sensing observations, which is important for space weather forcasting. We examine coronagraph observations in Section \ref{The Sun} and heliospheric consequences in Section \ref{HelioConseq}. The results are summarized and discussed in Section \ref{summary}.

\section{CMEs at the Sun}\label{The Sun}

On 2005 May 6, a CME (CME1) was launched from NOAA AR 10758 (S07{\degr}E28{\degr}), associated with a C8.5 flare that peaked at about 17:05 UT. There is no clear interplanetary type II burst accompanying this CME. Another CME (CME2) erupted on 2005 May 13 from NOAA AR 10759 (N12{\degr}E11{\degr}), approximately a week after the eruption of CME1. It was associated with a long-duration M8.0 flare that peaked at around 16:57 UT.

\begin{figure}
\epsscale{0.99}
\plotone{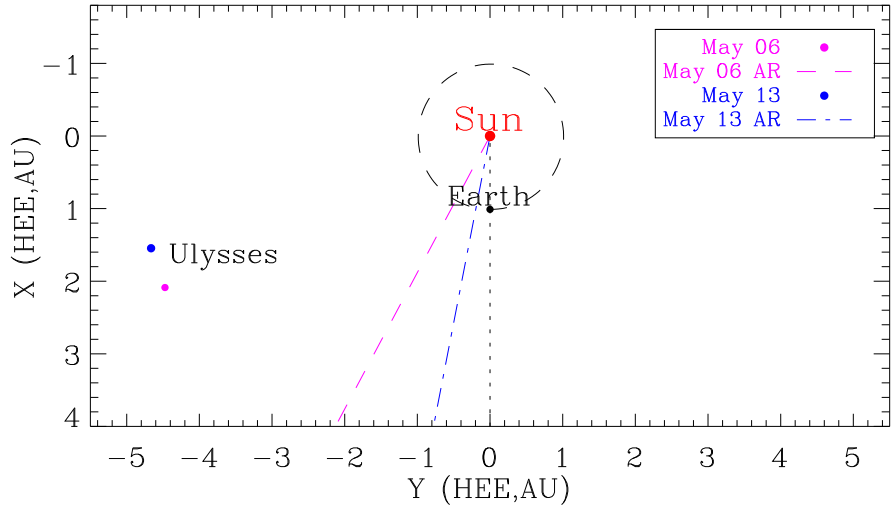}
\caption{\label{f1}Positions of \ulysses{} and the Earth in the ecliptic plane. The dashed curve indicates the orbit of the Earth. The purple dashed line and the purple dot represent the longitude of the active region and the position of \ulysses{} projected onto the ecliptic corresponding to the 2005 May 6 CME, respectively. The blue dot-dashed line and the blue dot indicate the longitude of the active region and the projected position of \ulysses{} corresponding to the 2005 May 13 CME, respectively.}
\end{figure}

Figure \ref{f1} shows the configuration of the planets and spacecraft in the ecliptic plane. The longitudes of the active regions (ARs) of CME1 and CME2 are also indicated in this figure. When the two CMEs erupted from the Sun, the longitudinal difference between the Earth and \ulysses{} for CME1 and CME2 was about 64.7{\degr} and 71.3{\degr}, respectively. Although the launch times of the two CMEs are well separated, they are likely to interact in  interplanetary space since the ARs were at similar longitudes with respect to the Earth/\ulysses{} and CME2 is faster than CME1 (see the text below). The CMEs would impact the Earth and perhaps \ulysses{} as well. 

\begin{figure}
\epsscale{0.8}
\plotone{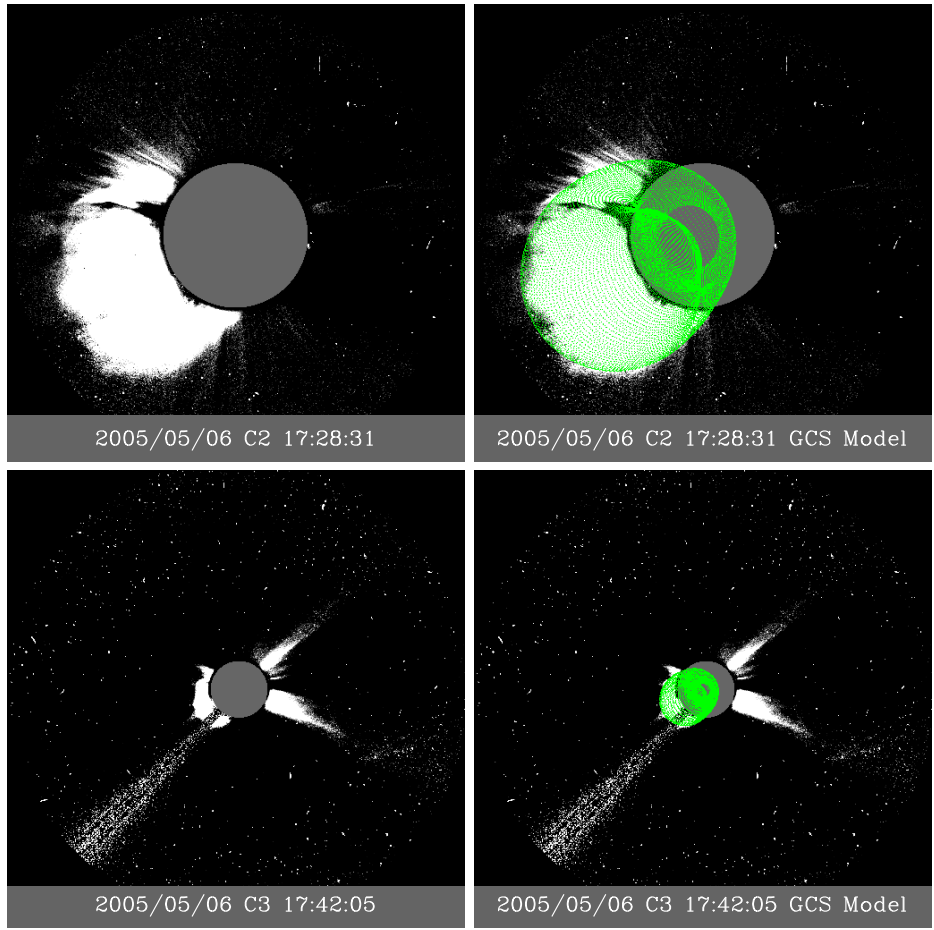}
\caption{\label{f2}Coronagraph running-difference observations and modeling of the 2005 May 6 CME. The left column shows the observations from LASCO C2 (upper) and C3 (lower), and the right column the GCS modeling (green grids) overplotted on the observed images. }
\end{figure}

The coronagraph observations and modeling of CME1 are displayed in Figure \ref{f2}. This is a partial halo CME, which is consistent with its solar source longitude (E28{\degr}). We use the GCS model proposed by \citet{2006Thernisien} to reproduce CME1 at different times based on the observations, which can give the heliocentric distances of CME1 leading front. This model has six free parameters: the longitude and latitude of the propagation direction, the tilt angle, aspect ratio and half angle of the flux rope, and the height of the CME leading front. While using this model, we keep the tilt angle, aspect ratio, and half angle parameters constant and adjust the longitude and latitude. We see a good visual consistency between the wireframe rendering obtained from the GCS model and the observed images. Applying the GCS model to single spacecraft observations may give rise to large uncertainties in the parameters. All the resulting parameters of the reconstruction are listed in Table \ref{tab:para}. The flux-rope tilt angle is chosen to be consistent with the orientation of the neutral line in the active region. The longitude of the propagation direction agrees with the solar source longitude (E28{\degr}). The latitude is also consistent with the solar source latitude (S07{\degr}), although a slight southward transition may be present. The speed of the CME, obtained with a linear fit of the CME leading front distances obtained from the GCS fit, is about 1300 \kms.

\begin{figure}
\epsscale{0.8}
\plotone{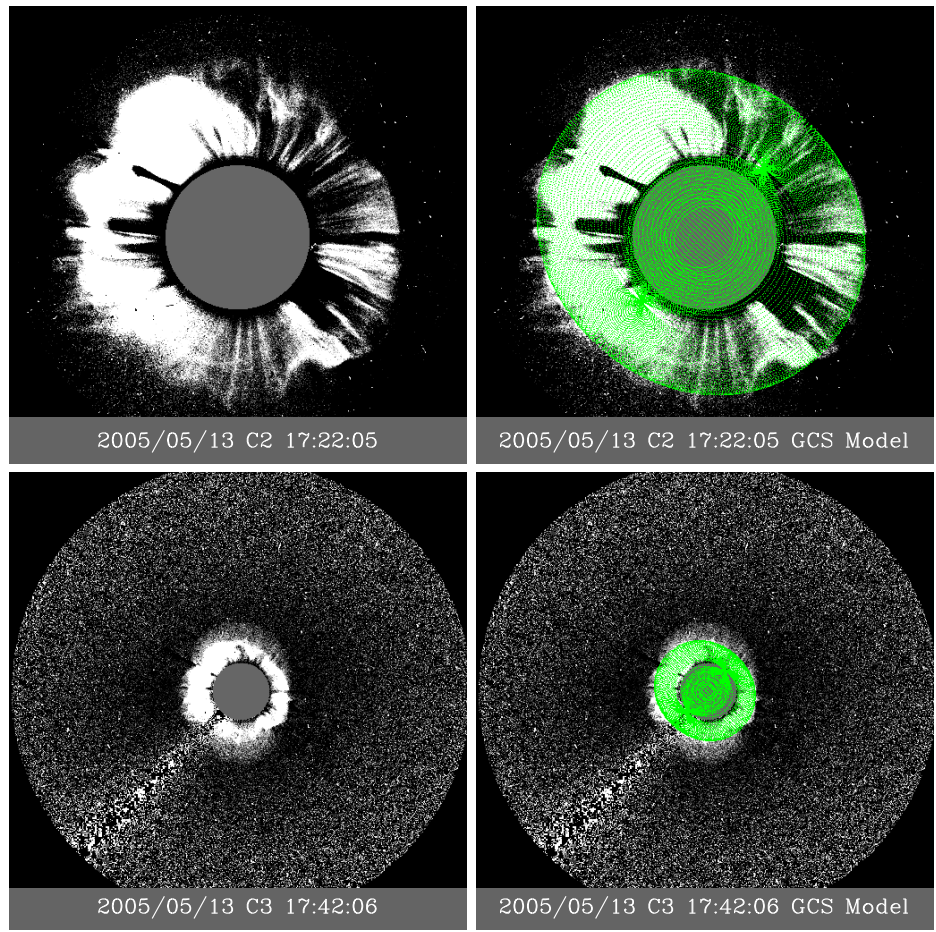}
\caption{\label{f3}Similar to Figure \ref{f2}, but for the coronagraph running-difference observations and modeling of the 2005 May 13 CME. }
\end{figure}

Figure \ref{f3} shows the coronagraph observations and modeling of CME2. This is a halo CME, and only two frames of images are available from LASCO. We also obtain a good visual agreement between the model results and the images of CME2. The tilt angle, again, is chosen to be consistent with the neutral line orientation in the active region inferred from magnetogram observations \citep{2006Yurchyshyn}. The longitude of the CME propagation direction also agrees with the solar source longitude (E11{\degr}). The latitude shows a small southward transition compared with the solar source latitude (N12{\degr}). The CME speed, estimated with a linear fit of the two distances, is about 2485 \kms. The error of the speed may be large, as we only have two data points.   

\section{Heliospheric consequences} \label{HelioConseq}

\subsection{ICMEs at the Earth and Geo-effectiveness} \label{1 AU ICME}

\begin{figure}
\epsscale{0.8}
\plotone{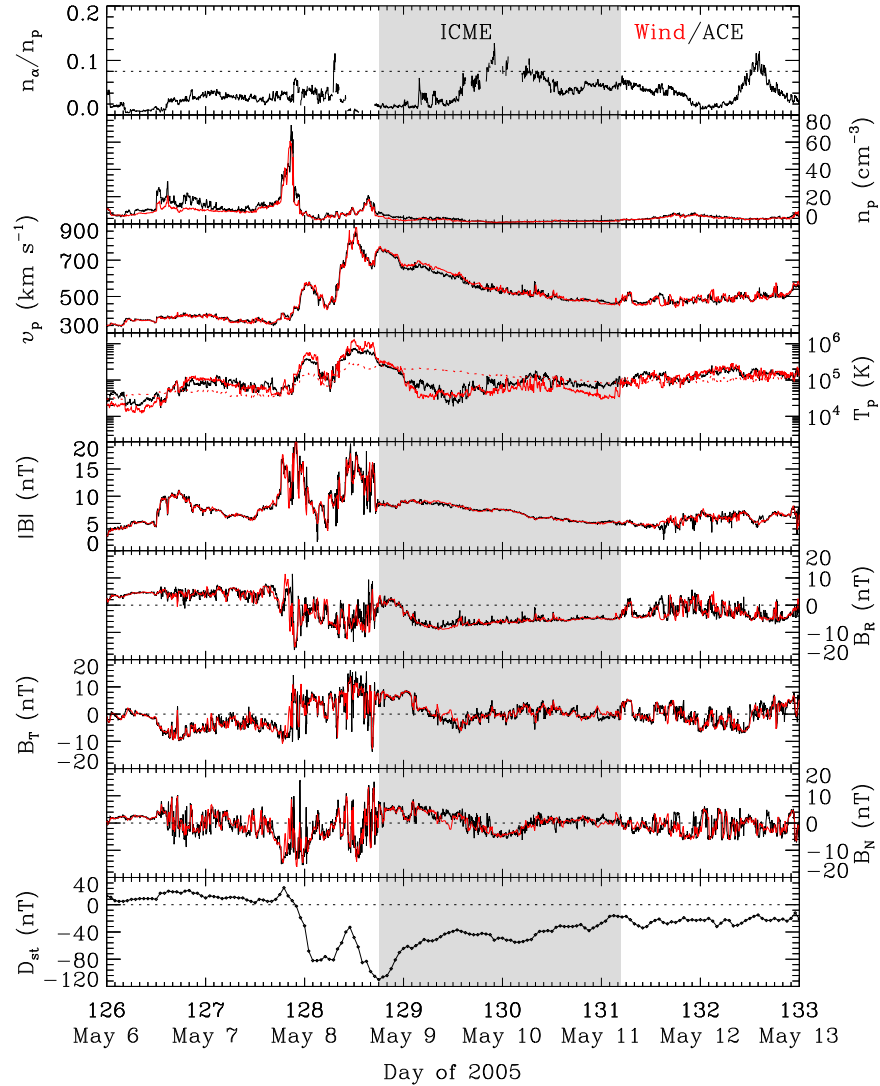}
\caption{\label{f4}Solar wind measurements of CME1 at \wind{} (red) and \ace{} (black). From top to bottom, the panels show the density ratio between alphas and protons, proton density, bulk speed, proton temperature, magnetic field strength and components, and D$_\mathrm{st}$ index, respectively. The shaded region indicates the ICME interval. The dotted horizontal line in the first panel marks the 0.08 level of the alpha-to-proton density ratio, which is often used as a threshold for ICMEs \citep{2005LiuRiBel, 2006LiuR, 2004Richardson}. The dotted curve in the fourth panel denotes the expected proton temperature from the speed observed by \wind.}
\end{figure}

We associate the two CMEs with their interplanetary counterparts at 1 AU based on their transit times from the Sun to the Earth. Figure \ref{f4} presents the \insitu{} measurements of CME1 by \wind{} and \ace{} near the Earth. There are no clear shock signatures ahead of the ICME. The shaded region indicates the ICME with an interval from 18:00 UT on May 8 to 04:33 UT on May 11. The ICME interval is mainly determined from the smooth magnetic field and the depressed proton temperature compared with the expected temperature. The use of the expected temperature for the identification of ICMEs is introduced by \citet{1993Richardson} and \citet{1995RichardsonC} based on the solar wind temperature-speed relationship developed by \citet{1986Lopez}. The alpha-to-proton density ratio is enhanced inside the ICME but patchy. \citet{2007Zhang} suggest that the solar wind structure ahead of the ICME is a corotating interaction region (CIR). The average solar wind speed across the ICME leading edge is about 750 \kms. Through comparing the speed of the ICME leading edge with that measured by \soho{} (1300 \kms), we suggest that CME1 must have decelerated within 1 AU. An intense geomagnetic storm occured with a minimum value of the D$_\mathrm{st}$ index of ${-}$110 nT. The southward field component inside the ICME is small (see the second panel from the bottom in Figure \ref{f4}). The intense geomagnetic storm is induced by the southward field components in the CIR ahead of the ICME although fluctuating \citep{2007Zhang}. The ICME may have contributed to the geomagnetic storm by compressng the CIR from behind.

\begin{figure}
\epsscale{0.8}
\plotone{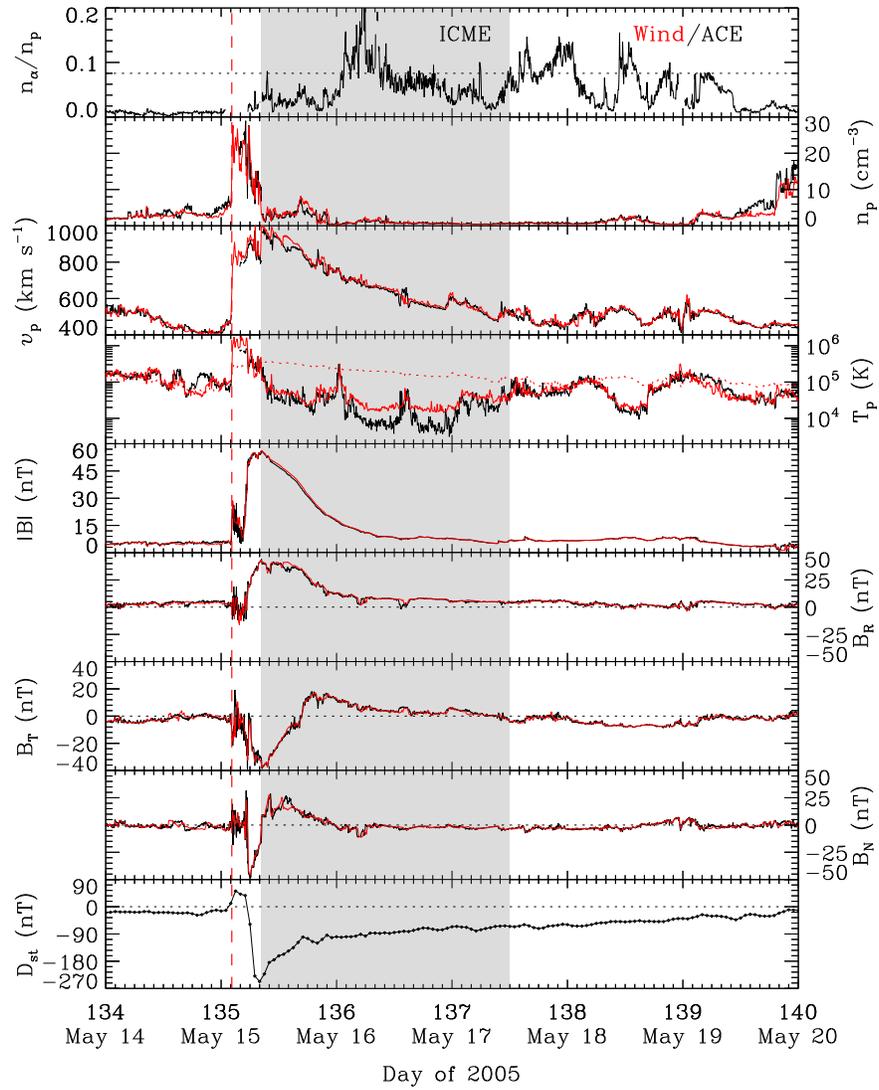}
\caption{\label{f5}Similar to Figure \ref{f4}, but for the \insitu\ measurements of CME2. The red vertical dashed line marks the arrival time of the shock.}
\end{figure}

Figure \ref{f5} shows \wind{} and \ace{} \insitu{} measurements of the 2005 May 13 CME (CME2). We can see a shock preceding the ICME at about 02:09 UT on May 15 inferred from the sharp increases in the solar wind plasma and magnetic field parameters. The position where the proton density, speed, temperature, and the magnetic field strength begin to decline indicates the ICME leading edge. It is difficult to determine the trailing boundary of the ICME interval, as the plasma and magnetic field parameters do not give a consensus on that. We tentatively suggest 12:00 UT on May 17 for the trailing boundary based on the depressed proton temperature, but 03:36 UT on May 19 is also possible considering the sustained period of the enhanced alpha-to-proton density ratio although the interval seems too long in this case. The transit time of CME2 from its eruption to the arrival of the shock at 1 AU is about 33.5 hr, which gives an average speed of about 1240 \kms. We calculate the shock speed near the Earth with a simplified equation of conservation of mass:
\begin{equation}\label{eq1}
v_{s}=\frac{n_{2}v_{2}-n_{1}v_{1}}{n_{2}-n_{1}}
\end{equation}
where $n_{1}$ and $n_{2}$ are the avarage densities and $v_{1}$ and $v_{2}$ are the average speeds upstream and downstream of the shock, respectively. Comparison between the average transit speed of CME2 ($\sim$1240 \kms), the LASCO speed ($\sim$2485 \kms) and the shock speed near the Earth ($\sim$950 \kms) suggests that the shock must undergo a deceleration when it propagates in the Sun-Earth space. A very severe geomagnetic storm with a minimum value of the D$_\mathrm{st}$ index of ${-}$247 nT, seems caused by the enhanced southward magnetic field in the sheath region between the shock and ICME in combination with the high speed there. Given the uncertainty in determining the ICME leading edge, the southward field component causing the geomagnetic storm may be partly from inside the ICME. The southward field component is as large as ${-}$47 nT in this case.

\subsection{ICMEs at \ulysses}\label{ulysses}

\begin{figure}
\epsscale{0.8}
\plotone{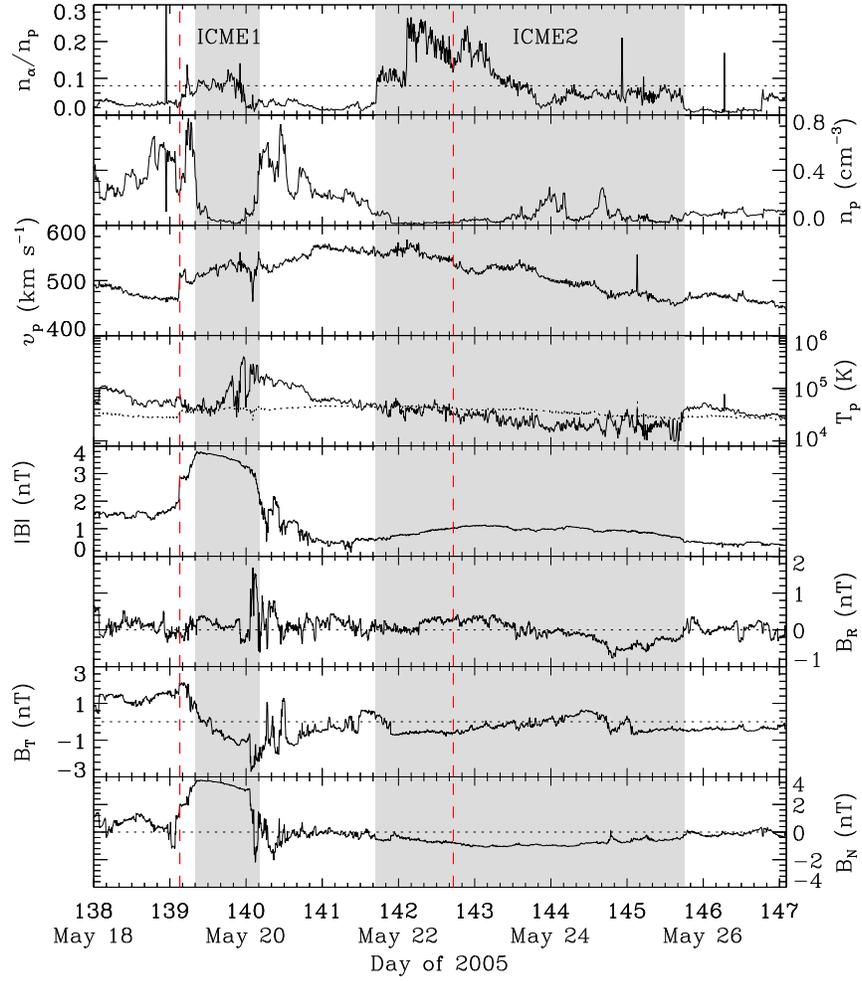}
\caption{\label{f6}Solar wind measurements at \ulysses. Similar to Figure \ref{f4}. The dotted curve in the fourth panel is the expected proton temperature from the observed speed by taking into account the temperature-distance gradient \citep[\textit{R}$^{-0.7}$;][]{2005LiuRiBel}. The first red vertical dashed line marks the arrival time of a shock-like structure, and the second one marks the arrival time of the MHD model-predicted shock for CME2.}
\end{figure}

Ulysses was at a distance of $\sim$5.08 AU from the Sun and $\sim$$21.4^{\circ}$ south and $\sim$$80^{\circ}$ east of the Earth, when it observed the 2005 May 6 and 13 CMEs. The positions of \ulysses{} projected onto the ecliptic plane corresponding to the two events are plotted on Figure \ref{f1}. Ulysses may observe both of the two CMEs given that the longitudes of the ARs are not far away from \ulysses. Figure \ref{f6} displays solar wind measurements by \ulysses. We identify two ICMEs as shown by the shaded regions from May 19 to May 26, which agree with the intervals of magnetic clouds (MCs) given by \citet{2010Du} and \citet{2014Richardson}. The first ICME (ICME1) is mainly determined from the enhanced magnetic field and rotation of the field, which is likely the \insitu{} counterpart of the 2005 May 6 CME (CME1) according to the transit speed (see the text below). The signatures of the magnetic field components of ICME1 at \wind{} and \ulysses{} are not the same, perhaps due to the longitudinal separation between the two spacecraft so that they observe different parts of it. It is also possible that the CME was modified during transit from 1 AU to \ulysses. There is a shock-like structure preceding ICME1. The transit time of ICME1 from the Earth to \ulysses{} is about 10 days, which implies an average speed of about 708 \kms. The average solar wind speed across the ICME1 leading edge is about 504 \kms. Comparison between the near-Earth speed ($\sim$750 \kms), the transit speed ($\sim$708 \kms) and the speed at \ulysses{} ($\sim$504 \kms) indicates that ICME1 underwent a gradual deceleration when it propagated from the Earth to \ulysses. The major deceleration occured from the Sun to 1 AU when we compare these speeds with the LASCO speed near the Sun ($\sim$1300 \kms).

We identify the second ICME (ICME2) by combining the increased alpha-to-proton density ratio, depressed proton temperature and smooth magnetic field. ICME2 is likely the \insitu{} counterpart of the 2005 May 13 CME (CME2). The CME2-driven shock disappeared at \ulysses. The average speed across the ICME2 leading edge is about 552 \kms. Again, the major deceleration of the CME occured during transit from the Sun to 1 AU by comparing the LASCO speed near the Sun ($\sim$2485 \kms), the shock speed near the Earth ($\sim$950 \kms) and the speed at \ulysses{} ($\sim$552 \kms), and thereafter ICME2 experienced a gradual deceleration. There is likely an interaction between the two ICMEs at \ulysses{} as the density and temperature are enhanced in the region between the two ICMEs and the speed of the ICME2 leading edge is larger than that of ICME1. The second ICME may thus have had an extra deceleration by interacting with the first one. However, it is difficult to know whether the two ICMEs have merged or not at \ulysses{} given the one-dimensional nature of the \insitu{} measurements.  

\subsection{CME/shock interplanetary propagation}\label{propa}

\begin{figure}
\epsscale{0.8}
\plotone{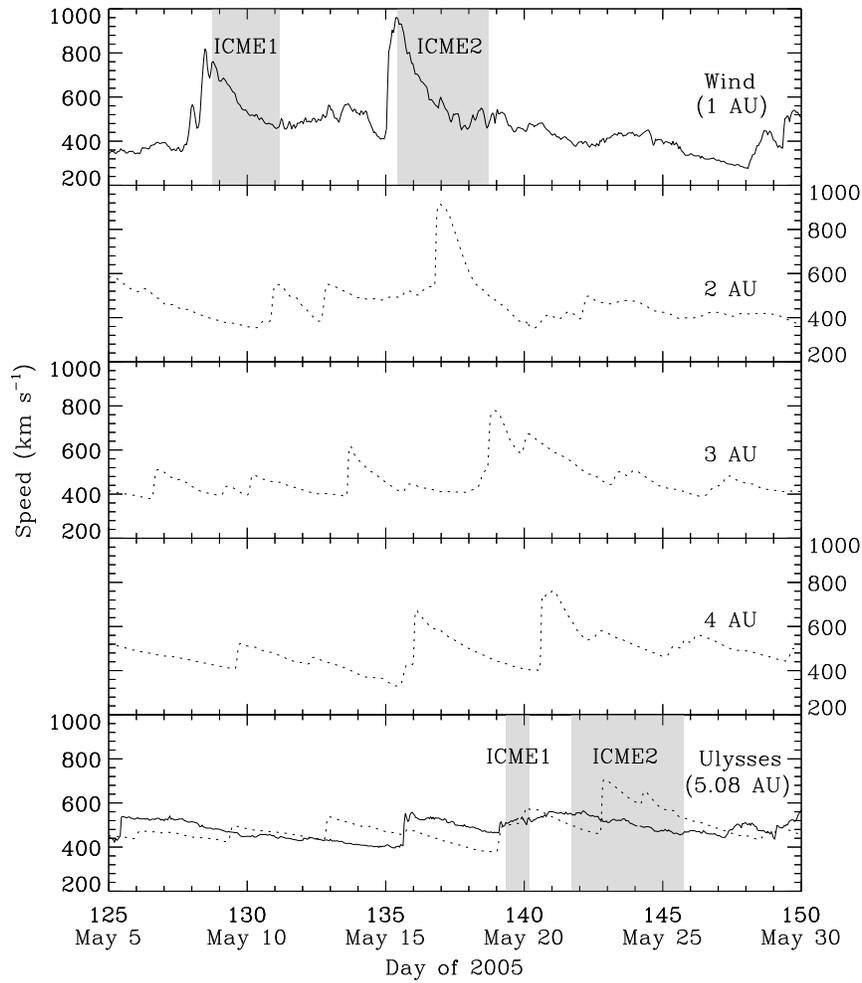}
\caption{\label{f7}Evolution of solar wind speeds from \wind{} to \ulysses{} via the 1-D MHD model. All of the panels have the same scales. The solid curves show the solar wind speeds at \wind{} and \ulysses, and the dotted curves denote the predicted speeds at 2, 3, 4, and 5.08 AU. The shaded regions represent the observed ICME intervals at \wind{} and \ulysses. }
\end{figure}

We use a 1-D MHD model \citep{2000Wang} with the solar wind disturbances observed at \wind{} as input, in order to show whether the observations at the Earth and \ulysses{} are consistent with the propagation of the ICMEs. This MHD model assumes spherical symmetry since the solar wind \insitu{} measurements are one-dimensional \citep[e.g.,][]{2001Wang, 2002Richardson, 2005Richardson, 2006Richardson, 2006Liu, 2008LiuL, 2011Liu, 2007Du}. Figure \ref{f7} shows the solar wind speed output by the MHD model at certain distances, and the observed ones at \wind{} and \ulysses, respectively. The streams associated with ICMEs are still persistent during transit from 1 AU to \ulysses. A shock forms from the stream associated with ICME1 and is coincident with the shock-like structure observed at \ulysses. The predicted arrival time at \ulysses{} is only about 0.7 hr earlier than observed. The time difference ($\sim$0.7 hr) is much smaller than the propagation time from \wind{} to \ulysses{} ($\sim$240 hr), so the shock-like structure is well tracked by the MHD model. The predicted arrival time of the ICME2 shock at \ulysses{} is  about 19:12 UT on May 22, which is about 26 hr later than that of the ICME2 leading edge. Note that the ICME2 shock is missed from \ulysses{} measurements. The model stream shows an enhanced speed compared with the observed speed within ICME2 at \ulysses, but is generally consistent with the ICME2 interval at \ulysses. In summary, the model gives a better agreement for ICME1 than for ICME2, which is probably due to the smaller longitudinal separation between the Earth and \ulysses{} during the time of CME1 (Figure \ref{f1}).

\begin{figure}
\epsscale{0.9}
\plotone{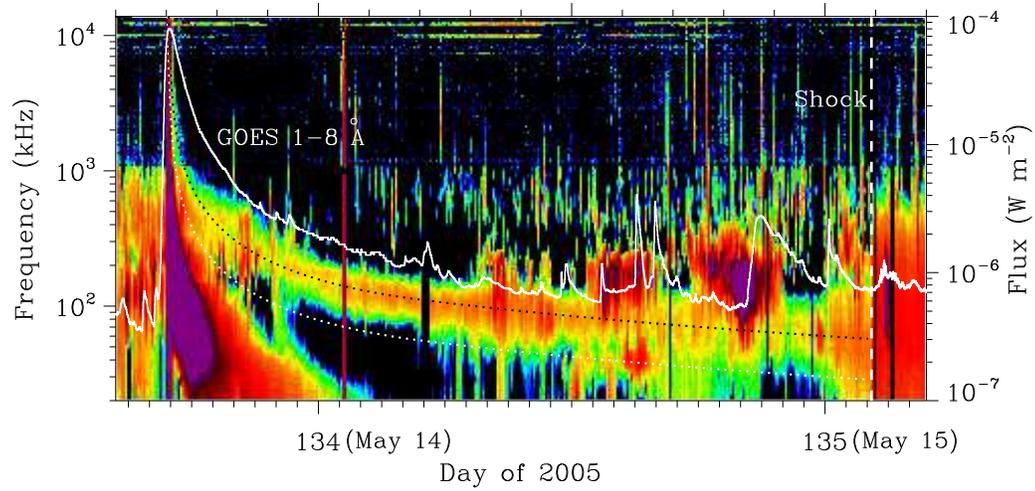}
\caption{\label{f8}Dynamic spectrum (colors) associated with the 2005 May 13 CME from \wind{} WAVES and flare X-ray flux (white curve) from \goes. The vertical dashed line indicates the arrival time of the shock at the Earth. The black dotted curve represents the best fit of the type II band at the second harmonic plasma frequency, and the white dotted curve is obtained by dividing the best fit by a factor of 2. }
\end{figure}

Type II radio bursts can be used to track a shock from the Sun out to 1 AU and fill the vacancy of direct solar wind plasma observations \citep[e.g.,][]{2007Reiner, 2008LiuL}. Figure \ref{f8} provides an overview of the radio dynamic spectrum associated with the 2005 May 13 CME (CME2). No clear type II burst was observed associated with the 2005 May 6 CME (CME1). An intense complex type III radio burst started at about 16:45 UT on May 13, which was almost coincident with the peak of the soft X-ray flux. \citet{1973Lin} suggest that a type III radio burst is caused by electron beams escaping from the flaring region. Such an intense type III burst is a sign of a major CME on the Sun \citep{2001Reiner}. Type II radio bands at the fundamental plasma frequency were observed from about 17:00 UT to 18:50 UT on May 13. After about 19:00 UT on May 13, diffuse type II radio bursts were mainly observed at the harmonic plasma frequency until the arrival of the shock at 1 AU on May 15. The type II burst supports our association between the May 13 CME and the ICME at 1 AU.

The plasma frequency, $f_{p}\left(\rm{kHz}\right)=8.97\sqrt{n\left(\rm{cm{^{-3}}}\right)}$, can be converted to heliocentric distance $r$ (in units of AU) through the Leblanc density model \citep{1998Leblanc}: 
\begin{equation}\label{eq2}
n=\frac{n_{0}}{7.2}\left(ar^{-2}+br^{-4}+cr^{-6}\right)\rm{cm^{-3}}
\end{equation}
where $a=7.2, b=1.95\times 10^{-3}, c=8.1\times 10^{-7}$, and ${n}_{0}$ is the electron density near the Earth in units of cm$^{-3}$. Combining the frequency drift of the type II radio burst and the shock parameters at the Earth, we can obtain a height-time profile of the shock within 1 AU. We adopt a kinematic model similar to the approach of \citet{2008LiuL}, \citet{2001GopalswamyL} and \citet{2007Reiner}. The model assumes that the shock has an initial speed ${v}_{0}$ near the Sun, a constant deceleration ${a}$ with a deceleration time period ${t}_{1}$, and a subsequent constant speed ${v}_{s}$. It does not include an acceleration phase, so its deceleration profile can be considered as an average over the acceleration and rapid deceleration stages in \citet{2013Liu}. The kinematic model of the shock can be expressed as
\begin{equation}\label{eq:r}
r=\begin{cases}
d+v_{s}\left(t-t_{T}\right)+a\left(\frac{1}{2}t^{2}+\frac{1}{2} t_{1}^{2}-t_{1}t\right), & \textrm{$t < t_{1},$}\\ 
d+v_{s}\left(t-t_{T}\right), & \textrm{$ t \geq t_{1},$}
\end{cases}
\end{equation}
where $d=1$ AU and ${v}_{0}={v}_{s}-{a}{t}_{1}$. The initial speed ${v}_{0}$ is determined from LASCO observations ($\sim$2485 \kms). The shock speed ${v}_{s}$ and the transit time ${t}_{T}$ can be calculated from the \insitu{} measurements at \wind. The model has two free parameters (${n_{0}}$ and ${a}$). We select the harmonic plasma frequencies of the type II radio burst and then fit the frequencies using the kinematic model. We adjust ${n}_{0}$ for the fit and find that a value of ${n}_{0}= 10$ cm$^{-3}$ can yield a best fit which simultaneously matches the type II band and the shock parameters at 1 AU. Two curves at the fundamental and harmonic plasma frequencies are obtained from the best fit, as shown in Figure \ref{f8}. The fit is consistent with the type II radio band at the harmonic frequency very well. The fit gives a constant deceleration ${a} \approx -35.7$ m s$^{-2}$ and a deceleration time period of about 12.23 hr with a deceleration cessation distance of about 0.52 AU ($\sim$112 R$_\sun$) from the Sun.   

\begin{figure}
\epsscale{0.99}
\plotone{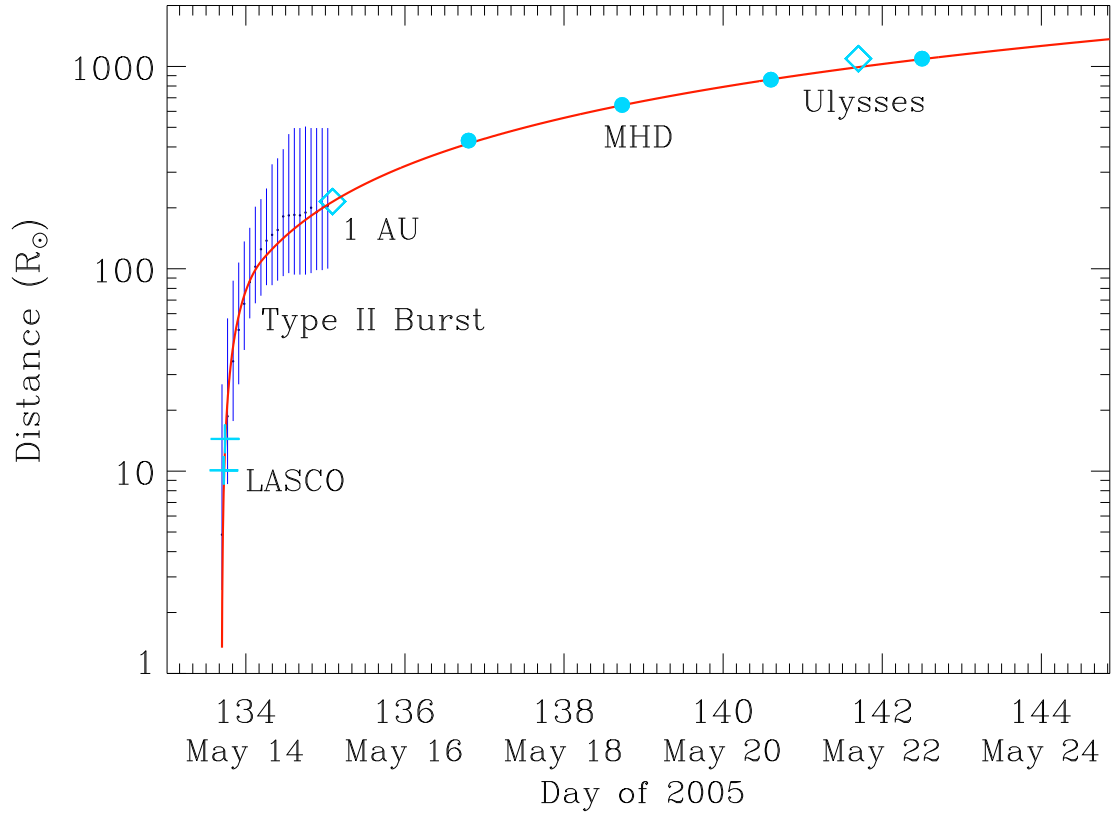}
\caption{\label{f9}Height-time profile (red solid curve) of the CME2-driven shock determined from the frequency drift of the type II radio burst and 1 AU shock parameters (where R$_\sun$ is the solar radius). The two crosses are the distances obtained from the GCS model based on the LASCO coronagraph observations. The small black dots with blue error bars indicate the distances obtained from the type II radio burst based on the Leblanc density model. The error bars are acquired according to the bandwidth of the type II radio burst. The two diamonds show the distances of \wind{} and \ulysses{} when the shock and the ICME2 leading edge arrived at them, respectively. The large blue dots denote the shock arrival times at 2, 3, 4, and 5.08 AU (\ulysses) predicted by the MHD model.}
\end{figure}

Figure \ref{f9} shows the distances obtained from the GCS modeling of the LASCO images, the frequency drift of type II radio burst, \insitu{} measurements at \wind{} and \ulysses{} and MHD model output. We extend the fit in order to show whether it matches the distances beyond 1 AU. The fit is consistent with the GCS distances near the Sun and MHD model output outside 1 AU, although it is obtained from only the type II burst and 1 AU shock parameters. The curve is a little lower than the \ulysses{} location but not significantly. Again, this is likely due to the longitudinal and latitudinal separations between the Earth and \ulysses, possible different solar wind backgrounds and changes in the CME/shock structure. This height-time curve is generally consistent with the rapid deceleration and then gradual deceleration phases found by \citet{2013Liu} for fast CMEs (${>}$ 1000 \kms). The deceleration cessation distance ($\sim$112 R$_\sun$) is larger than that of \citet[][$\leq$ 80 R$_\sun$]{2013Liu}, but this depends on the coronal and solar wind background and may thus vary. 

\section{summary and discussion}\label{summary}
   
We have analysed the evolution of two geo-effective CMEs in the heliosphere, which occurred on 2005 May 6 and 13, combining \soho{} coronagraph observations, frequency drift of the type II radio burst measured by \wind{} WAVES (only for the second CME), and \insitu{} measurements at \wind, \ace{} and \ulysses{} with modeling techniques. This work provides insights into the propagation characteristics of CMEs from the Sun far into interplanetary space as well as space weather forcasting. The results are summarized and discussed as follows.    

Comparision between the CME speed near the Sun, the speed at the Earth and finally the speed at \ulysses{} ($\sim$5.08 AU) indicates that both of the CMEs had their major deceleration take place inside 1 AU and thereafter underwent a gradual deceleration. This is consistent with the general picture of the Sun-to-Earth propagation profile of fast CMEs discovered by \citet{2013Liu} using \stereo{} wide-angle imaging observations, i.e., a rapid deceleration before reaching 1 AU followed by a gradual deceleration or a roughly constant speed (note that we did not include the acceleration phase here due to limitation of the present onservations). Based on the type II radio burst and the 1 AU shock parameters, we obtain a best fit of the height-time curve for the 2005 May 13 CME (CME2) and find that the rapid deceleration of CME2 ceased at about 112 R$_\sun$. This gives further evidence for the rapid deceleration of fast CMEs before reaching 1 AU, although the cessation distance of the rapid deceleration is somewhat larger than those in \citet{2013Liu}. The best fit also matches CME distances near the Sun and MHD model output at different distances, and is roughly consistent with the arrival of the ICME2 leading edge at \ulysses. This indicates the application of a simple kinematic model to a large range of distances (from the Sun to about 5 AU), which is important for space weather forcasting. Comparision of the kinematics between the two CMEs suggests that the faster the CME the larger the deceleration within 1 AU. As the deceleration is due to interactions with the ambient solar wind, this result confirms that CMEs tend to comove with the ambient solar wind before reaching 1 AU \citep{1999Lindsay, 2001GopalswamyL, 2016Liu}.

The results also suggest that the two CMEs may start to interact around 5 AU at \ulysses, although their eruption times on the Sun were well separated (by about a week). There is no evindence that the two CMEs interacted with each other near the Earth, as can be seen from the \insitu{} measurements at \wind{} and \ace. When the second CME occured, the first CME had already passed beyond 1 AU. They may interact with each other far into interplanetary space due to the the fact that the ARs were at similar longitudes with respect to the Earth/\ulysses{} and CME2 is faster than CME1. The result that the two CMEs interacted in the deep interplanetary space indicates the ``fate'' of CMEs in the heliosphere, i.e., different CMEs may eventually merge and form merged interaction regions (MIRs) in the outer heliosphere with their identities lost \citep[e.g.,][]{1984Burlaga, 2002Richardson, 2014LiuR}. 

Two intense geomagnetic storms occured with minimum D$_\mathrm{st}$ value of ${-}$110 nT and ${-}$247 nT, respectively. For the first case without a shock, the geomagnetic storm was induced by the southward field components in the CIR ahead of the ICME \citep{2007Zhang}. The ICME may have contributed to the geomagnetic storm by compressing the CIR from behind. For the second case with a preceding shock, the geomagnetic storm was primarily caused by the southward field components between the shock and ejecta. Both cases stress the critical role of the sheath region in geomagnetic storm generation. 

\acknowledgments
The research was supported by the Recruitment Program of Global Experts of China, the NSFC under grant 41374173, and the Specialized Research Fund for State Key Laboratories of China. We acknowledge the use of data from \textit{SOHO}, \textit{GOES}, \wind, \ace, \ulysses, and the D$_\mathrm{st}$ index from WDC in Kyoto. 

\clearpage

\begin{deluxetable}{ccccccccc}
\tablecaption{CME Parameters Obtained from the GCS Model. \label{tab:para}}
\tablehead{
\colhead{CMEs} & \colhead{{Coronagraph}} & \colhead{Time} & \colhead{Lon($^{\circ}$)\tablenotemark{a}} 
& \colhead{Lat($^{\circ}$)\tablenotemark{a}} & 
\colhead{Tilt Angle} & \colhead{Aspect Ratio} & 
\colhead{Half Angle} & \colhead{{Height (R$_\sun$)}\tablenotemark{b}}
}
\startdata
       & C2 & 17:28 & -35.78 & -15.65 & 35.78 & 0.674 & 15.93  & 6.58  \\
       & C2 & 17:54 & -32.42 & -15.65 & 35.78 & 0.674 & 15.93  & 9.65  \\
       & C3 & 17:42 & -28 & -14.54 & 35.78 & 0.674 & 15.93  & 8.39  \\
       & C3 & 18:18 & -28 & -14.54 & 35.78 & 0.674 & 15.93  & 13.16  \\
  CME1 & C3 & 18:42 & -28 & -16.77 & 35.78 & 0.674 & 15.93  & 15.77  \\
       & C3 & 19:42 & -28 & -19.01 & 35.78 & 0.674 & 15.93  & 22.71  \\
       & C3 & 20:18 & -28 & -22.36 & 35.78 & 0.674 & 15.93  & 25.05  \\
       & C3 & 20:42 & -28 & -22.36 & 35.78 & 0.674 & 15.93  & 28.56  \\
\hline
  CME2 & C2 & 17:22 & -1.12 &  -1.12 & -43.04 & 0.696 & 9.78  & 10.10  \\
       & C3 & 17:42 & -2.24 &  -1.68 & -43.04 & 0.696 & 9.78  & 14.42  
\enddata
\tablenotetext{a}{Longitude and latitude in the heliographic coordinates. }
\tablenotetext{b}{The heliocentric distances of CME leading front at different times.}
\end{deluxetable}

\clearpage
\bibliography{references}
\bibliographystyle{aasjournal}
\end{document}